\documentclass[prb,showpacs,preprint]{revtex4-1}%
\usepackage{amsfonts}
\usepackage{amsmath}
\usepackage{amssymb}
\usepackage[dvips]{graphicx}%
\usepackage{float}
\usepackage{epsfig}
\usepackage{subfigure}
\usepackage{multirow}
\begin{document}

\title {Nitrogen-induced Ferromagnetism in BaO}

\author{Gul Rahman}\email{gulrahman@qau.edu.pk}

\address{Department of Physics,
Quaid-i-Azam University, Islamabad 45320, Pakistan}

\begin{abstract}
Density functional theory with local spin density approximation has been used to propose possible room temperature ferromagnetism in N-doped NaCl-type BaO. Pristine BaO is  a wide bandgap semiconductor, however, N induces a large density of states at the Fermi level in the nonmagnetic state, which suggests magnetic instability within the Stoner mean field model. The spin-polarized calculations show that N-doped BaO is a true half- metal, where N has a large magnetic moment, which is mainly localized around the N atoms and a small polarization at the O sites is also observed. The origin of magnetism is linked to the electronic structure. The ferromagnetic(FM) and antiferromagnetic (AFM) coupling between the N atoms in BaO reveal that doping N atoms have a FM ground state, and the calculated  transition temperature ($T_{C}$), within the Heisenberg  mean field theory, theorizes possible room temperature FM in N-doped BaO. Nitrogen also induces ferromagnetism when doping occurs at surface O site and has a smaller defect formation energy than the bulk N-doped BaO. The magnetism of N-doped BaO is also compared with Co-doped BaO, and we believe that N has a greater potential for tuning magnetism in BaO than Co. 
\end{abstract}

\maketitle

\section{Introduction}
\label{sec:intro}
Inducing magnetism in non-magnetic oxides is one of the main areas of interest in the research community. Generally, many oxides are insulator and, without magnetic impurity atoms, non-magnetic. However, recent theoretical and experimental research of oxides has show ferromagnetism with defects or light elements, e.g., C, N, Li.\cite{apl,jmmm,gr6,gr7,Mg,01,02,ZnOprl,tio2-2011} In the past decade, density functional theory (DFT) has played a vital role in either proposing new magnetic materials or elucidating the origins of defects-driven magnetism in oxides.\cite{1a, KCa, Nit, Mg,zn, gr6, gr7, grprb,12,13} 
To date, there are a few oxides that  show room temperature (RT) ferromagnetism, e.g., ZnO, SnO$_{2}$, In$_{2}$O$_{3}$, CeO$_{2}$.\cite{zn,8,9,10} Light elements can not only be used to develop magnetism in oxides, but can also be used to stabilize intrinsic defects in the host materials.\cite{arxi,znoc} Through the combined efforts of theoreticians and experimentalists, we are now certain that  magnetism in oxides induced by non-magnetic impurities.~\cite{hf,hf2,01,02} 
Now it is known that magnetism develops in the
nonmagnetic oxides where the impurity atoms have a finite local magnetic
moment and these local magnetic moments will interact with each other to form a net magnetic moment in the host material. Those materials in which magnetism is induced by doping nonmagnetic impurities, for which the substitutional atom can have a finite magnetic
moment and the $2p$-electrons of the doped atom plays an essential role in governing 
magnetism in the host material, are generally considered as $d^{0}$ magnetic materials.
These nonmagnetic impurity atoms can also form impurity
bands in the bandgap of host material, and ferromagnetic behavior can be
expected if the Fermi energy lies within these impurity bands. 

BaO is an oxide with interesting structural and electronic
properties. It can be used as a NO$_{3}$ storage device
for catalysis.\cite{bao1} BaO is also considered as a
precursor to the well known ferroelectric perovskite oxide BaTiO$_{3}$, which can have either TiO$_{2}$ or BaO terminated surface when BaTiO$_{3}$ is grown on a suitable substrate as a thin film. Possible magnetism in TiO$_{2}$.\cite{02} has been extensively studied, however, less attention has been given to impurities in BaO and there are no detailed theoretical and experimental studies on inducing magnetism in it. It has been experimentally
observed that bulk BaO  naturally occurs in a B1 (NaCl) structure.\cite{bao-exp} Recently, experimental reports have claimed the growth of ultrathin BaO films on SrTiO$_{3}$(001) substrate, and the formation of BaO nanoparticles on reconstructed SrTiO$_{3}$(001), while a locally ordered c($4\times\,4$) BaO structure is observed on the disordered sample surface.\cite{exp-2013} Tan \textit{et.al}, \cite{tan-2011} also observed RT ferromagnetism in N-doped BaTiO$_{3}$ and the origin of the magnetism was correlated 
with the presence of N. Therefore,
studying the electronic properties of BaO is
 of great importance for the  development of a new applications, specially in the area of magnetism.
Hence, we propose to add a new functionality, i.e. magnetism, in BaO. We show that N in BaO has a ferromagnetic (FM) ground state and the transition temperature $T_{\rm{C}}$, which is the temperature at which a material goes from a paramagnetic(disorder phase)state to a magnetic phase(order phase) state, is above  room temperature.

\section{Computational Methods}
\label{sec:reslt}
To study the magnetism of N-doped BaO, we performed calculations in the framework of density functional theory, ~\cite{DFT} using linear combination of atomic orbital basis as implemented in the SIESTA code ~\cite{siesta}. A double-$\zeta$ polarized basis set for all atoms was used. The local spin density approximation~\cite{lda}(LSDA) was adopted for describing exchange-correlation interactions. We used standard norm-conserving pseudopotentials ~\cite{ps} in their fully nonlocal form ~\cite{pss}.  A cutoff energy of 400 Ry for the real-space grid was adopted. The sampling of $k$-space was performed with Monkhorst and Pack (MP) scheme with a regularly spaced mesh of  $18\times 18\times 18$. Convergence with respect to $k$-point sampling and cutoff energy was carefully checked.

To investigate the magnetism and electronic structures of N-doped BaO, we considered the $2\times2\times1$ (Ba$_{4}$O$_{3}$N$_{1}$), $2\times2\times2$ (Ba$_{8}$O$_{7}$N$_{1}$), $2\times2\times3$(Ba$_{12}$O$_{11}$N$_{1}$) supercells of the primitive unit cell of NaCl-type BaO. {Calculations were also carried out using a $3\times3\times3$ (Ba$_{27}$O$_{26}$N$_{1}$) supercell.} In all these supercells, N was doped at O site due to smaller difference in their atomic sizes and electro-negativities. For the magnetic properties of BaO (001)surface, we considered the conventional unit cell of BaO, and used different $n\times\,m\times\,z$  symmetric slabs of thickness $z$, defined as a monolayer of BaO. We investigated $n=1,m=1,z=8$ (Surf$_{1}$) and $n=2,m=1,z=8$ (Surf$_{2}$) surfaces of BaO and doped N at O on both sides of the BaO surface. We added a vacuum region of about $\sim$ 10\;\AA, so that the two surfaces do not interact with each other through the vacuum region. 
Additional simulations, using plane waves plus pseudopotentials as implemented in the Quantum Espresso (QE) code\cite{qe1}, were also carried out to further test the validity of our results. 
Atomic positions were relaxed, using conjugate-gradient algorithm,~\cite{cg} until the residual Hellmann-Feynman force on single atom converges to less then 0.05 eV/\AA. {Test calculations, using LSDA$+U$, were also carried out by considering the on-site Coulomb correction ($U= 6.0$ eV, our previously optimized value\cite{arxi}) between the $p$-orbital electrons of O.\cite{ldau1,ldau2,arxi}}

\section{Results and Discussions}
\label{sec:reslt}

\begin{figure}\hspace{-3.55cm}
\includegraphics [width=0.8\textwidth]{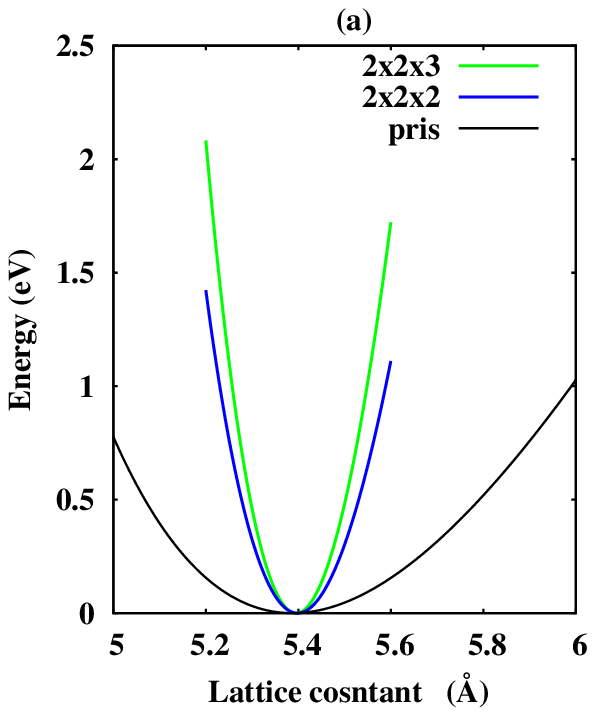}\hspace{-7.cm}
\includegraphics [width=0.6\textwidth]{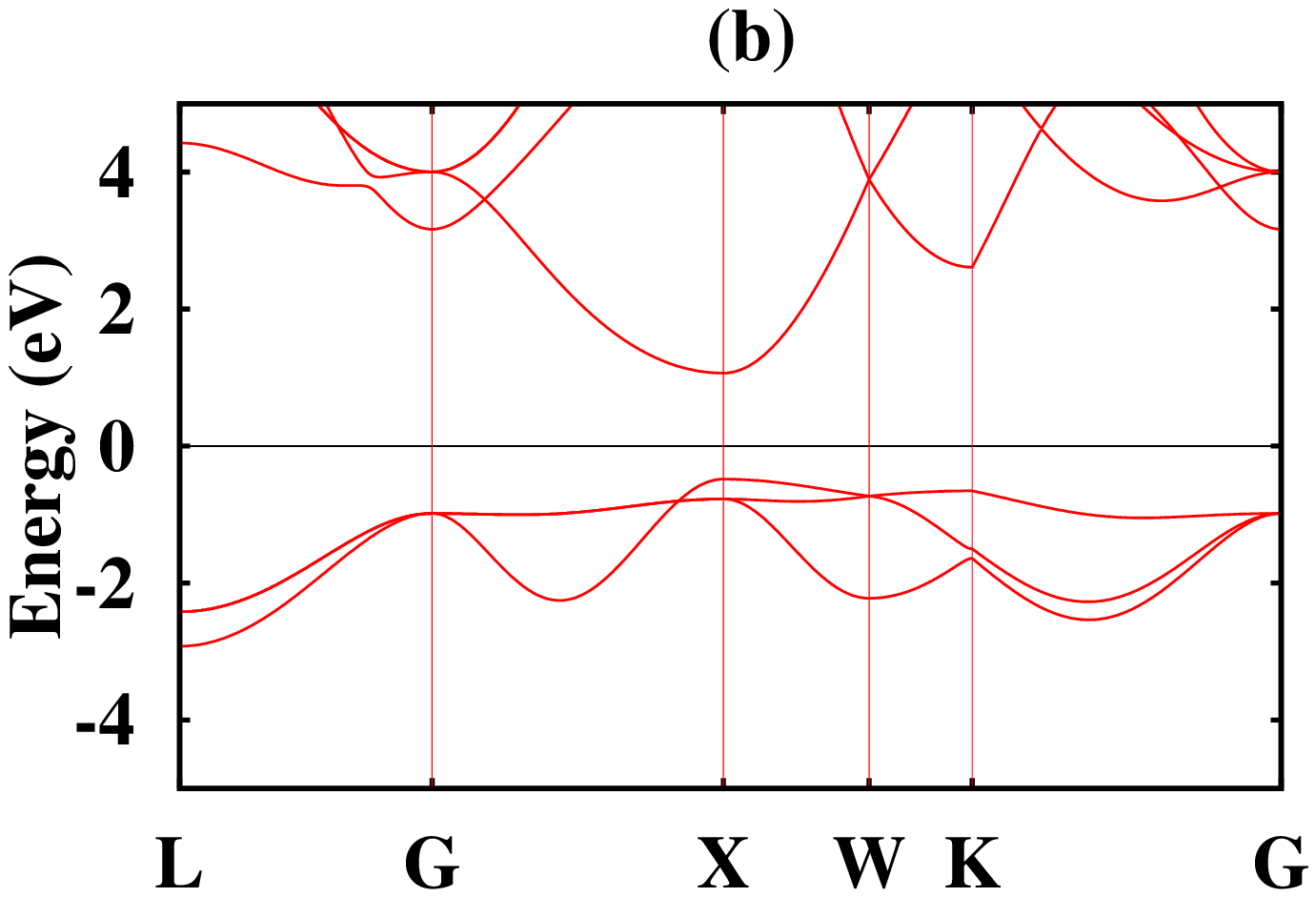}\hspace{-0.3cm}
\caption{
(a) The calculated lattice constant vs energy of pristine BaO (black line), Ba$_{8}$O$_{7}$N$_{1}$ (blue line), and Ba$_{12}$O$_{11}$N$_{1}$( green line). (b) 
 The band structure of pristine BaO, and the horizontal line (black) shows the Fermi energy, which is set to zero. }
\label{Band_BaO}
\end{figure}

\begin{figure}
\hspace{-1.0cm}
\includegraphics [width=0.65\textwidth]{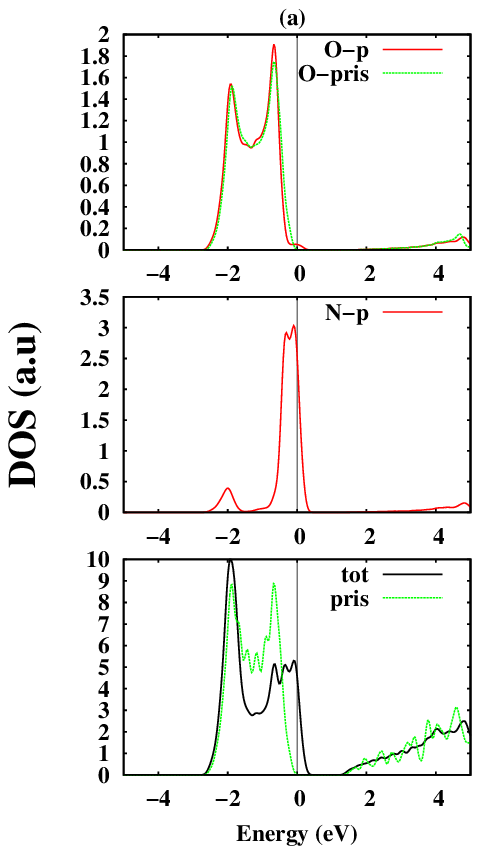}\hspace{-7.5cm}
\includegraphics [width=.65\textwidth]{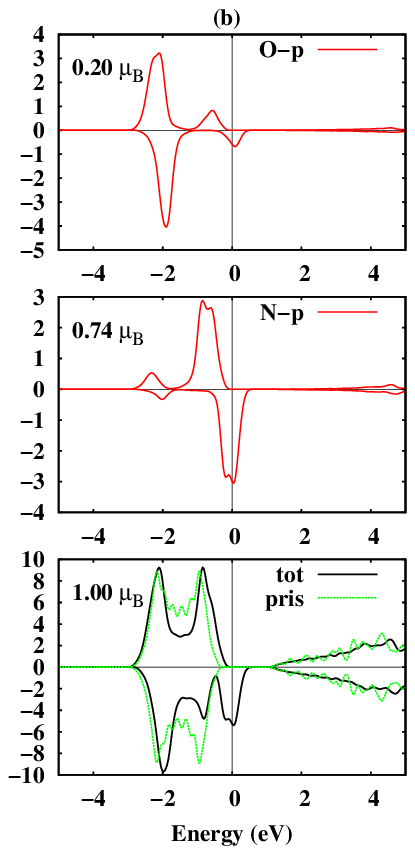}\hspace{-7.5 cm}
\includegraphics [width=.65\textwidth]{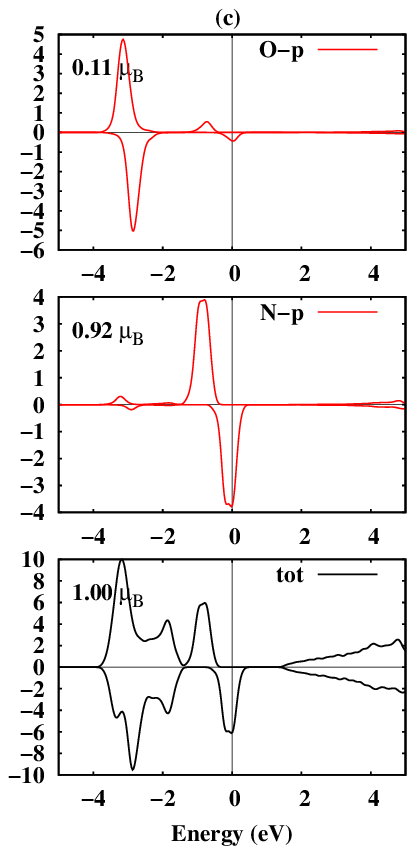}\hspace{-7.5 cm}
\includegraphics [width=.65\textwidth]{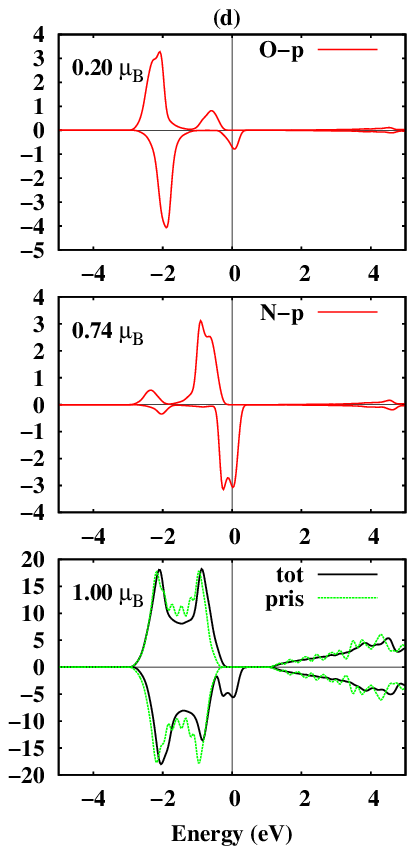}\hspace{-7.5 cm}
\includegraphics [width=.65\textwidth]{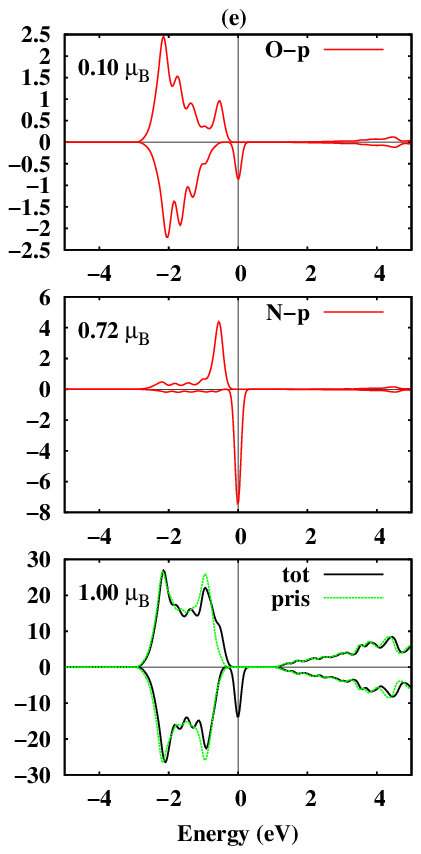}\hspace{-2.5 cm}
\caption{
The calculated total and atom projected density of states (PDOS) of Ba$_{4}$O$_{3}$N$_{1}$ in the non-magnetic (a), where the green dotted lines show the PDOS(O-pris) and total DOS (pris) of pristine BaO. The spin-polarized (magnetic)
total and PDOS of Ba$_{4}$O$_{3}$N$_{1}$, Ba$_{8}$O$_{7}$N$_{1}$, and Ba$_{12}$O$_{11}$N$_{1}$ are shown in (b), (d), and (e), respectively. Panel (c) shows LSDA$+U$ DOS of Ba$_{4}$O$_{3}$N$_{1}$. The green dotted lines in the total DOS panels show the total DOS of the pristine BaO (pris) of the same supercells. The total magnetic moments per N atom and local magnetic moments of N and O are also shown. The O and N PDOS are represented by solid (red) lines. The vertical lines represent  the Fermi energy, which is set to zero. }
\label{FM-DOS}
\end{figure}

First, we calculated the optimized lattice constant of NaCl-type BaO, which was found to be $5.40$\AA\,, as shown in Fig.~\ref{Band_BaO}(a). The calculated lattice constant is comparable to the previous calculated $5.47$\AA,\cite{baohf1} and experimental $5.52$\AA\cite{latbao} values. Our DFT estimated value is smaller than the experimental value due to the underestimation problem of DFT-LDA.
Using the optimized lattice constant, we calculated the band structure, which is shown in Fig.~\ref{Band_BaO}(b). We see that BaO is a wide bandgap semiconductor and the calculated bandgap at the $X$ point is $\sim 1.62 $ eV. The band is degenerate at $G$ point [$2\pi/a(0,0,0)$], but this degeneracy is removed at $X$ point where the band has less curvature (large effective mass)  compared with point $G$. The electronic density of states (DOS) shows that bands near the Fermi energy ($E_{\rm{F}}$) are mainly derived from the O $p$-orbitals [see Fig.~\ref{FM-DOS}(a)]. 
{The LSDA calculated indirect ($\Gamma$-$X$)bandgap is $\sim 2.0$ eV, in agreement with the previous LDA calculation.\cite{gap-LDA} However, the experimental and GW calculated value is 3.88 eV~\cite{gap-exp,gap-exp2,gap-LDA}. We then used LSDA$+U$, and our LSDA$+U$ calculated bandgap is $\sim3.6$, which is comparable with the GW calculated and experimental values. }

{It is important to consider atomic relaxation in N-doped BaO. Usually, the doping impurity at the host site can either compress or elongate the bond length depending on the atomic size of the impurity atom. In some cases the impurity atom can also be located at an interstitial site and  form a defect complex.\cite{arxi} We therefore relaxed all the atomic coordinates of BaO:N and no significant changes in the atomic position of N at O sites were found. To further confirm this, both Ba$_{8}$O$_{7}$N$_{1}$  and Ba$_{12}$O$_{11}$N$_{1}$ were considered, and calculated their optimized lattice constants, which are $~5.40$\AA\, for both the systems [see Fig.~\ref{Band_BaO}(a)]. This lattice constant is approximately equal to our optimized lattice constant of pristine BaO, which shows that N can easily  dope the O sites without any structural distortion.}


Figure~\ref{FM-DOS}(a) shows the calculated total and atom projected (P) DOS of N-doped BaO in the non-magnetic state. For comparison purposes, the total PDOS of pristine BaO is also shown. It can be clearly seen that the DOS of doped and pristine BaO are the same in the conduction and valance bands. However, near the Fermi energy ($E_{\rm F}$) the DOS of the doped system is drastically changed and the material shows metallic behavior, in contrast to pristine BaO, which is an insulator. The PDOS demonstrates that the main peak at the Fermi energy arises mainly from contributions from the $p$-orbitals of N. The N atom in BaO forms an impurity band in the bandgap of BaO. A small impurity-derived peak can also be seen at 2.0 eV just below $E_{\rm{F}}$, and this peak hybridizes with the $p$-orbitals of O atoms. Therefore, the O and N atoms are only hybridize near the Fermi energy. No significant changes in the PDOS of O  can be seen when the  O sites are doped with N, except a small shoulder, derived by the N $p$-orbitals, near the Fermi energy. The calculated results show that the
top of the valence band consists of N $2p$ electrons, which are higher than O $2p$ electrons. The large DOS at the Fermi energy $D(E_{F})$ in the N-doped BaO shows that there is an instability towards magnetism within the Stoner mean field theory of magnetism.\cite{Stoner} Within the Stoner model, which was mainly proposed for itinerant electron system, the large $D(E_{F})$ can lead to a large Pauli susceptibility which  is large enough for the band to split spontaneously, and magnetism in N-doped BaO can be expected.

To confirm that N-doped BaO can present magnetism within the Stoner mean field theory, we further carried out spin-polarized calculations. The spin-polarized electronic density of state of Ba$_{4}$O$_{3}$N$_{1}$ is shown in  Fig.\ref{FM-DOS}(b). The total DOS clearly shows that N induces magnetism in BaO when doped at O sites. The spin-polarized structure shows that the non-magnetic impurity band in the bandgap [see Fig.\ref{FM-DOS}(a)] is mainly due to the  contribution from  minority spins. Large exchange splitting at N site can also be seen, as expected in the Stoner mean field theory. 
For comparison, we also show the total DOS of pristine BaO and one can easily judge that the conduction band of Ba$_{4}$O$_{3}$N$_{1}$ in the spin-up state is identical to that of BaO. 
The $2p$
states of nitrogen are located at the top of the valence band, which are mainly derived from the anion $p$-states, resulting in the hybridization of electron wave functions of the $2p$ orbitals. The majority
spin-states of nitrogen also hybridize with oxygen $2p$ states, promoting the
the impurity band which connects to the top
of the valence band. Such chemical process rise  to a bandgap in the majority spins states. The minority spins on the other hand create an impurity band
in the bandgap which includes $E_{\rm{F}}$. As a result this
impurity band is broadened having about 0.38 eV half-width, and the approximated exchange splitting is 0.80 eV. The majority spins are completely occupied and behaving just like an insulator, whereas the minority spins are partially occupied and have a metallic nature. Such electronic structure is the fingerprint of a half-metal ferromagnet,~\cite{dgroot,grprb1} which has applications in the area of spin electronics. 
The total and local magnetic moments of Ba$_{4}$O$_{3}$N$_{1}$ were also calculated, and the calculated total magnetic moment is $1.0\,\mu_{\rm{B}}$ per unit cell. The PDOS clearly shows that the magnetic moment arises  mainly from contributions from the N $p$-orbitals and the local magnetic moment at the N site is $\sim 0.74\, \mu_{\rm{B}}$. A small induced magnetic moment ($\sim 0.20\, \mu_{\rm{B}}$) at the nearby O site is also observed. The PDOS of O shows some unoccupied minority spin states, which were occupied in the pristine BaO, suggesting that the holes induced by N also localized at the O $p$-orbitals. 
We repeated the same calculations by doping N at O site in $2\times2\times2$ and $2\times2\times3$ supercells, and interestingly we found half-metallic behavior in all these doped systems [see Fig.~\ref{FM-DOS}(d and e)] and the magnetic moments were mainly localized at the N sites. 
It is also noticeable that N does not polarize the whole valance band, it polarizes only the valance electronic states near the Fermi energy. Such behavior of N in BaO is different from that of transition metals in oxides/semiconductors, which usually polarize the whole valance band.\cite{dms} The band-width of  N  decreases and the band is narrowed, as the concentration of N is reduced in BaO [see Fig.~\ref{FM-DOS}(e)], which further increases the localization of $2p$ states of N. This localization of the N $2p$ states may further increase the observation/suppression of ferromagnetism/antiferromagnetism in BaO, which is discussed in the following paragraph. 
The total magnetic moment per N atom is  $1.0\,\mu_{\rm{B}}$, which   indicates that the total magnetic moment is independent of the N concentrations. In all these N-doped BaO systems, the      
calculated magnetic moment is consistent with Hund's rules indicating that the N
dopant exists as a N$^{2-}$ ($s^{2}p^{5}$) anion illustrating that each N impurity
introduces one $2p$ hole and the strong Hund-type exchange makes the nitrogen $2p$ spin-up orbitals
completely occupied, and the spin-down states are partially filled [Fig.\ref{FM-DOS}(b-e)]. As
a result, one can expect the (ferro)magnetism in N-doped BaO to be stabilized mainly by the predominant double-exchange mechanism.\cite{Stoner} Before we address the question of FM or antiferromagnetic (AFM) coupling between the N atoms, the above calculations were also {performed for Ba$_{27}$O$_{26}$N$_{1}$ and Ba$_{32}$O$_{31}$N$_{1}$, and we confirmed} that N in BaO has spin-polarized band structure. We also confirmed the above conclusions using the QE code. It is also important to comment on the magnetism of N-doped BaO using LSDA$+U$. The LSDA$+U$ calculated total and PDOS of Ba$_{4}$O$_{3}$N$_{1}$ are shown in Fig.~\ref{FM-DOS}(c). As expected, the N forms an impurity band in the bandgap of BaO and N-driven oxygen $p$ states can also be seen in the gap region of BaO. Similar observations have  also been observed in self-interaction correction (SIC) for C-doped BaO.\cite{BaC} The local magnetic moment of N (O)is increased (decreased), but the total magnetic moment of the unit cell remains unchanged, i.e., $1.0\mu_{{\rm B}}$. Similar conclusions were also drawn when LSDA$+U$ calculations were performed for Ba$_{8}$O$_{7}$N$_{1}$, and Ba$_{12}$O$_{11}$N$_{1}$ consistent with our previous work.~\cite{arxi}

{As predicted, N can induce half-metallic magnetism in insulating BaO. However, there remains a question on whether N-doped BaO can have FM order above room temperature, and whether  the exchange
coupling sufficiently large to  give rise to possible RT ferromagnetism. Magnetism is a cooperative phenomenon and a single N in BaO can not determine the true magnetic ground state. Therefore,  we consider $2\times2\times3$ and { $3\times3\times3$ supercells} of BaO and doped two N atoms at different O sites. The distance between the two N atoms $d$ was varied, as it was the FM and AFM coupling between them. {For comparison purposes, we kept the same N-N separation in both  supercells, and the results are summarized in Table~\ref{table}}. It is enchanting to see that both BaO:N systems have FM ground state, and the total magnetic moment of the unit cell is $2.0\,\mu_{\rm{B}}$ (i.e., $1.0\,\mu_{\rm{B}}$ per N). All the systems remained half-metallic as  well.  The strength of the exchange interactions $J$ can be judged  from $\Delta E=E_{AFM}-E_{FM}$, where $E_{AFM}$($E_{FM}$) is the total energy of the supercell in the AFM (FM) state. This exchange energy can be further used to estimate $T_{C}$. Using the Heisenberg mean field model ($k_{\rm{B}}\,T_{C}=2\,\Delta\,E$/3),\cite{TC1,TC2} the estimated $T_{C}$ is found to be close to room temperature, indicating  the possibility of room temperature ferromagnetism in BaO:N systems. {One can also see that both the systems have different transition temperatures due to different N concentrations in BaO.}
Note that, usually, mean field theory overestimates $T_{C}$ but it can give an indication of possible RT FM. Therefore, RT ferromagnetism in N-doped BaO is expected, if properly prepared. It is encouraging to mention that our estimated $T_{C}$  matches with N-doped BaTiO$_{3}$~\cite{tan-2011} and other magnetic oxides~\cite{tio2-2011,znoc} indicating that the magnetism is mainly attributed to N impurities.

The formation enthalpy $H_{f}$ of BaO is calculated using $H_{f}=E(\rm{BaO})-[E(\rm{Ba})+\frac{1}{2}E(\rm{O}_{2})]$, where $E(\rm{BaO})$, $E(\rm{Ba})$, and $E(\rm{O}_{2})$ are the total energies of NaCl-type BaO, BCC Ba, and oxygen molecule, respectively. The calculated $H_{f}$ was found to  
be $-6.49$ eV, which comparable with the previous calculated (-$5.64$\cite{baohf1}, -$5.19$\cite{baohf2}), and experimental (-$5.74$\cite{baohf3}) values.
{We followed our previous approach~\cite{arxi} and calculated the defect formation energy $E_{f}$ of N in BaO 
under  Ba-rich and O-rich conditions because the defect formation energy strongly depends on the growth condition and chemical potential. \cite{arxi}  The calculated $E_{f}$ of 
Ba$_{4}$O$_{3}$N$_{1}$, Ba$_{8}$O$_{7}$N$_{1}$, and Ba$_{12}$O$_{11}$N$_{1}$ 
in Ba-rich (O-rich) conditions are -1.15 (5.33), -1.14 (5.34), and -1.15 (5.45) eV, respectively. It is clear to see that N in BaO has a negative defect formation energy under Ba-rich condition, which further increases the possibility of introducing  N in BaO.
{Note that LSDA$+U$ calculated $E_{f}$ of Ba$_{4}$O$_{3}$N$_{1}$ is about 6.6 eV in O-rich condition, which is increased by including $U$ parameter--similar behvior has  also been  observed in other oxides.\cite{arxi} }
We then varied the distance $d$ between the two N atoms, using  $2\times\,2\times\,3$ and $3\times\,3\times\,3$ supercells, and calculated $E_{f}$ under different growth conditions. The calculated $E_{f}$ per N atom of both  systems in the FM states is shown in Table~\ref{table}. The calculated $E_{f}$ values are in the range of a single N doped BaO, and therefore we can not expect clustering of N in BaO when doping the  O sites. The formation energy depends on the N concentration in BaO. We may safely conclude that all the studied systems have negative formation energies under  Ba-rich environments.  Such calculated results can help experimentalists to dope N in BaO.} The role of intrinsic defects (vacancies) can not be ignored when a sample is  synthesised, these intrinsic defects may interact with the impurity atom, and such interaction may further decrease $E_{f}$ of the impurity atom.\cite{arxi} There is also an experimental report on N-induced ferromagnetism in BaTiO$_{3}$, \cite{tan-2011} which suggests the possibility of  N-doping  at the O sites in BaO. Experimental work will be very helpful to further analyse the possibilities of N-doped  BaO.}

\begin{figure}
\hspace{-1cm}
\includegraphics [width=.75\textwidth]{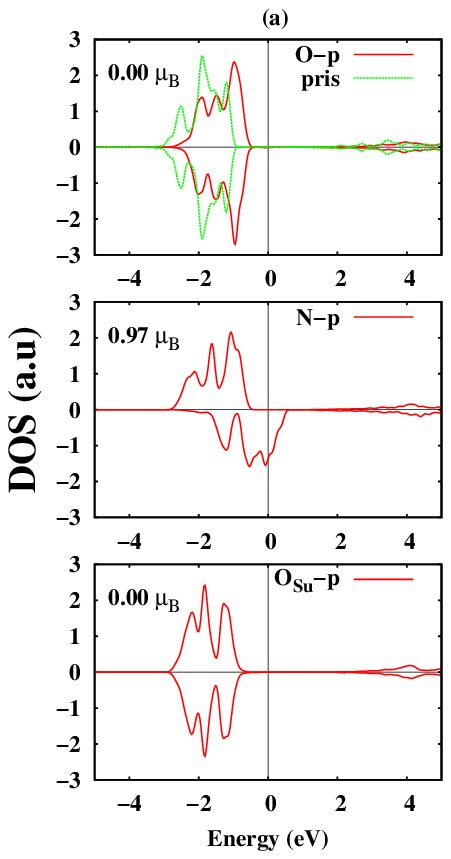}\hspace{-7.5cm}
\includegraphics [width=.75\textwidth]{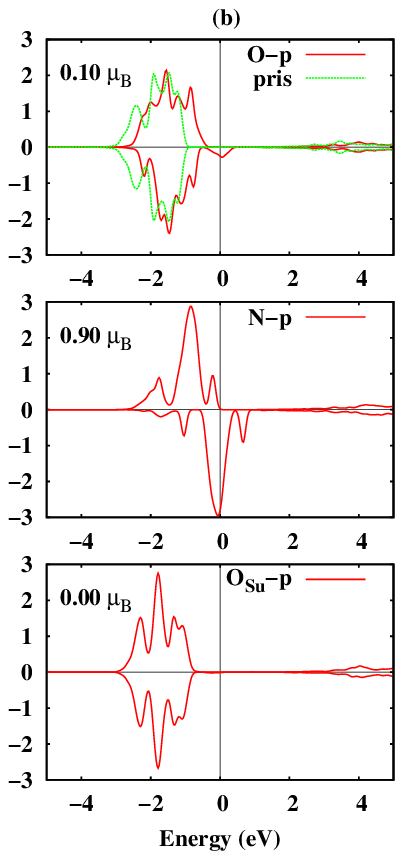}\hspace{-5.5cm}
\includegraphics [width=.2\textwidth]{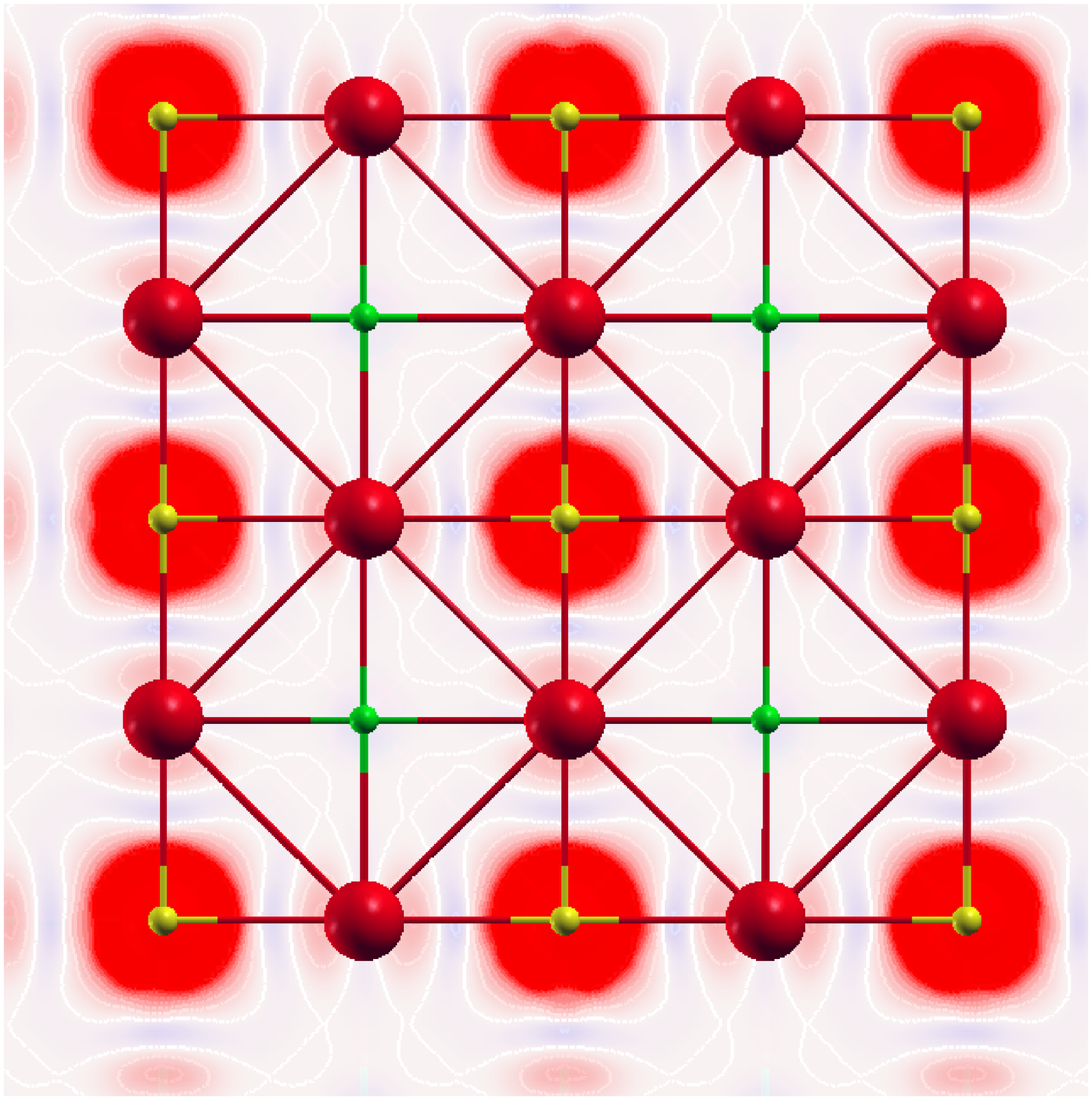}
\includegraphics [width=.1\textwidth]{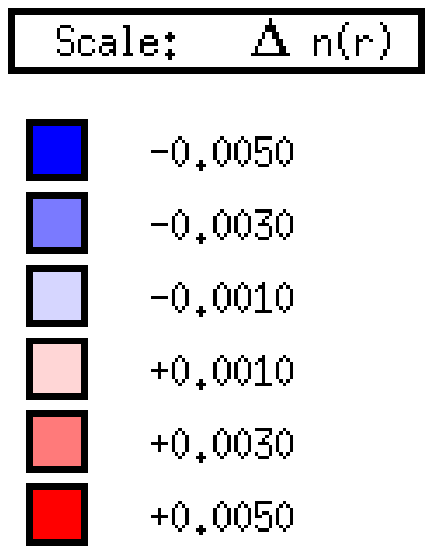}\hspace{-1.5cm}
\caption{The LSDA calculated surface atomic projected DOS of Surf$_{1}$(a) and Surf$_{2}$(b). The top panels show the PDOS of surface O atoms (red lines) along with the PDOS of clean surface O atoms (green lines).
The middle panels show the PDOS of surface N atoms (red lines).
The bottom panels show the PDOS of sub-surface O atoms denoted as O$_{\rm{Su}}$. 
The local magnetic moments of N and O atoms are also shown. The vertical lines show the Fermi energy, which is set to zero.
The calculated surface spin density of Surf$_{1}$(c), where red, green, and yellow balls represent Ba, O, and N atoms, respectively. The spin density of Surf$_{1}$ is calculated on a large surface to show zero polarization at surface O atoms. The inset shows the scale of the spin density.  }
\label{surface}
\end{figure}

To propose BaO for practical practical applications, e.g., in the area of thin films, it is  essential to study the electronic and magnetic properties of N doped at surface O site in BaO, Surf$_{1}$. We found that N can also induce magnetism in BaO (001) surface, and the magnetism is mainly localized at the N site as shown in Fig.\ref{surface}(a,b). Surface oxygen atoms have negligible magnetic moment and N has a large magnetic moment ($0.97 \mu_{\rm{B}})$ when doped at surface O sites, as expected for surface N atoms. No induced magnetic moment at the subsurface atoms were found, which is different from other doped oxides.\cite{apl,gr2014} The PDOS of surface O atoms are shifted towards the Fermi energy  compared to  clean surface O atoms, indicating that the band gap of BaO(001) surface is reduced due to surface states driven by N-O hybridization. 
The majority spins of N are occupied and the minority spins are partially occupied. The calculated surface spin densities of N, O, and Ba atoms are shown in Fig.\ref{surface}(c), and it is evident to see no spin-polarization at the surface O and Ba atoms is observed. The spin density is mainly localized around the N atoms. We then used $n\times\,m\times\,z$ surface (Surf$_{2}$) and doped two N atoms at two different O sites and considered the FM and AFM coupling between the N atoms. Here, we have also found that the FM coupling between the surface N atoms has lower energy than that of AFM coupling. As the concentration of N atom is increases, a small magnetic moments at the surface O atoms were also induced, but the sub-surface O atoms do not show any induced magnetic moment[Fig.\ref{surface}(b)]. No surface relaxation or reconstruction was observed when all the atoms were fully relaxed. {The calculated $E_{f}$ of N at surface O in Ba-rich (O-rich) condition is $-1.58$ ($4.90$) eV for Surf$_{1}$. 
For  Surf$_{2}$, when the two N atoms were doped at the surface O sites, the $E_{f}$ per N in Ba-rich (O-rich) conditions is $-1.48$ (5.00) eV when the distance between the two surface N atom is 3.82\AA, whereas when the separation between the two N atoms is 5.40\AA, the calculated $E_{f}$ per N in Ba-rich (O-rich) condition is $-1.60$ (4.88) eV. }
  Surface N-doped BaO has a smaller defect formation energy than the bulk N-doped BaO, similar to other oxides.\cite{gr2014} Note that a smaller surface $E_{f}$ corresponds to a larger N-N bond distance suggesting that N prefers isolation than clustering in BaO(001) surface.

We have shown that N in BaO can display possible room temperature ferromagnetism either in bulk or surface doped system, and finally we would like to compare it with the Co-doped BaO. Our preliminary results clearly indicate that a Co atom at either  O or Ba site induces magnetism with a larger(about 3 times) $E_{f}$ than BaO:N systems. Therefore, we  believe that N doping in BaO has the potential to easily tune magnetism and may be superior to Co doped BaO. Further experimental work would be needed to demonstrate the potential of BaO:N for possible applications in the area of spin electronics. We also believe that our prediction can also be helpful for BaO terminated surface of BaTiO$_{3}$.
\section{summary}
To summarize, DFT was used to investigate potential magnetism in N-doped NaCl-type BaO. Pristine BaO is known to be non-magnetic wide bandgap semiconductor, and N at O sites induce an impurity band in the bandgap of BaO. It has been  discussed that N in BaO can also induce magnetism and the magnetism arises  mainly from contributions by the N atoms. The minority impurity-driven band is narrowed with decreasing the N concentrations in BaO. The origin of magnetism was elucidated from the atom projected density of states. All the considered scenarios with different N concentrations in BaO showed half-metallic behavior. The ferromagnetic and antiferomagnetic interactions between the N atoms were also considered, and they revealed  that N-doped BaO is a ferromagnetic material. The transition temperature was also calculated using Heisenberg mean field model, and it  was found to be above room temperature. The magnetic interaction between the N atoms is limited to the next-nearest-neighbor interactions. The surface magnetism of N-doped BaO(001) was also studied by doping  surface O sites with N. N atoms at surface O sites also showed ferromagnetic ground state, and the magnetism was mainly limited to the BaO surface. No induced magnetic moments were observed at the sub-surface O sites. The defect formation energy showed that N has a smaller formation energy than the bulk N doped BaO. The magnetism of N-doped BaO was also compared to that  Co-doped BaO, and it was concluded N-doped BaO can easily be achieved. LSDA$+U$ calculations also supported ferromagnetism in N-doped BaO.

The author acknowledges insightful discussions with
{V\'{\i}ctor M. Garc\'{\i}a-Su\'arez} and the cluster facilities of NCP, Pakistan.


\newpage

\begin{table}
\caption{Calculated total magnetic moment $M$ (in $\mu_{\rm{B}}$), $\Delta E=E_{AFM}-E_{FM}$ per cell (in eV), transition temperature $T_{C}$ (in K), and defect formation energy $E_{f}$ per N (in eV) of N-doped BaO ($3\times\,3\times\,3$).  $Ba_{rich}$ and O$_{rich}$ shows  $E_{f}$ under Ba-rich and O-rich conditions, respectively.  The first column shows N--N separation $d$ (in \AA). Values in the brackets correspond to the $2\times\,2\times\,3$ supercell. }

\begin{tabular}{ cccccccccccccccccc }
\hline
&&$d$&&$M$&&$\Delta\,E$&&$T_{C}$&&&&\multicolumn{3}{ c }{$E_{f}$} \\
\hline
&&&&&&&&&&&& $Ba_{rich}$ && O$_{rich}$ \\ \hline
 &&3.82&&2.0(2.0)&&0.064(0.077)&&495(597)&&&& -1.05(-1.07)&&5.42(5.40)&&\\
&&6.61&&2.0(2.0)&&0.035(0.089)&&271(689)&&&&-1.02(-1.08)&&5.45(5.40)\\
\hline\hline
\end{tabular}
 \label{table}
\end{table}

\newpage

\newpage

\end{document}